\documentclass[
amsmath, %
floatfix, %
twocolumn, %
reprint, %
superscriptaddress, %
prb, %
aps, %
nofootinbib,%
nobibnotes,%
citeautoscript, %
lengthcheck, %
final, %
]{revtex4-1}
\renewcommand{\thispagestyle}[1]{}
\usepackage[T1]{fontenc}
\usepackage[utf8]{inputenc}
\usepackage{graphicx}
\usepackage{latexsym}
\usepackage{amsthm}
\usepackage[low-sup]{subdepth}

\usepackage{color}
\usepackage{mathtools}
\mathtoolsset{showonlyrefs}

\usepackage{placeins}

\usepackage[dvipsnames]{xcolor}

\usepackage[]{newtxtext}
\usepackage[subscriptcorrection,nosymbolsc,smallerops,bigdelims]{newtxmath}
\DeclareMathAlphabet{\mathcal}{OMS}{cmsy}{m}{n}
\DeclareMathAlphabet{\mathbcal}{OMS}{cmsy}{b}{n}
\usepackage{bm}

\newcommand{\overbar}[1]{\mkern 1.85mu\overline{\mkern-1.85mu#1\mkern-1.85mu}\mkern 1.85mu}
\newcommand{\widebar}{\overbar}

\usepackage{floatrow}
\usepackage{siunitx}
\usepackage{hyperref}
\hypersetup{
	colorlinks,
	linkcolor={blue!90!black},
	citecolor={blue!90!black},
	urlcolor=	{blue!90!black}
}

\let\oldeqref\eqref
\renewcommand*{\eqref}[1]{%
	\hyperref[#1]{\oldeqref{#1}}%
}

\newcommand{\mfrac}[2]{#1/#2}

\newcommand{\rr}{\bm{r}}

\newcommand{\phbr}{{\lambda}}
\newcommand{\phws}{q}
\newcommand{\phwv}{{\bm{\phws}}}

\newcommand{\ql}{{\phwv,\phbr}}

\newcommand{\hf}{\frac12}

\renewcommand\Re{\operatorname{Re}}

\newcommand{\kp}{{\bm{k} {\cdot} \bm{p}}}

\mathchardef\mhyphen="2D

\newcommand{\sumi}{\sum_{i}}
\newcommand{\sumij}{\sum_{ij}}

\newcommand{\sumijkl}{\sum_{ijkl}}

\newcommand{\ii}{\mathrm{i}}
\newcommand{\mr}[1]{\mathrm{#1}}
\newcommand{\e}{\mr{e}}

\newcommand{\oneu}{1}
\newcommand{\oned}{\widebar{1}}
\newcommand{\twou}{2}
\newcommand{\twod}{\widebar{2}}

\newcommand{\upa}{{\uparrow}}
\newcommand{\doa}{{\downarrow}}

\DeclarePairedDelimiter\lr{\lparen}{\rparen}
\DeclarePairedDelimiter\Lr{\lbrack}{\rbrack}
\DeclarePairedDelimiter\LR{\lbrace}{\rbrace}
\DeclarePairedDelimiter\abs{\lvert}{\rvert}

\DeclarePairedDelimiterX{\comm}[2]{\lbrack}{\rbrack}{#1, #2}
\DeclarePairedDelimiter\ket{\lvert}{\rangle}

\DeclarePairedDelimiterX{\braket}[2]{\langle}{\rangle}{#1\delimsize\vert #2}
\DeclarePairedDelimiterX{\ketbra}[2]{\rvert}{\lvert}{#1 \delimsize\rangle\!\delimsize\langle #2}
\DeclarePairedDelimiterX{\matrixel}[3]{\langle}{\rangle}{#1 \delimsize\vert #2 \delimsize\vert #3}

\newcommand{\hc}{\mr{H.c.}}

\makeatletter
\newcommand{\raisemath}[1]{\mathpalette{\raisem@th{#1}}}
\newcommand{\raisem@th}[3]{\raisebox{#1}{$#2#3$}}
\makeatother

\newcommand\Tstrut{\rule{0pt}{2.4ex}} 

\definecolor{cbred}{HTML}{e31a1c}
\definecolor{cbgreen}{HTML}{33a02c}
\definecolor{cbblue}{HTML}{176aa7}

\usepackage{dcolumn}
\newcolumntype{d}[1]{D{.}{.}{#1}}
\newcommand{\lab}[1]{\multicolumn{1}{c}{#1}}

\usepackage{xpatch}
\xpatchcmd\bibsection{19}{10}{}{}
\xpatchcmd\bibsection{\begingroup}{\vskip -15pt\begingroup}{}{}

\begin{document}
\title{Controllable electron spin dephasing due to phonon state distinguishability\\in a coupled quantum dot system}

\author{Micha{\l} Gawe{\l}czyk}
	\email{michal.gawelczyk@pwr.edu.pl}
	\affiliation{Department of Theoretical Physics, Faculty of Fundamental Problems of Technology, Wroc\l{}aw University of Science and Technology, 50-370 Wroc\l{}aw, Poland}
	\affiliation{Department of Experimental Physics, Faculty of Fundamental Problems of Technology, Wroc\l{}aw University of Science and Technology, 50-370 Wroc\l{}aw, Poland}

\author{Mateusz Krzykowski}
	\affiliation{Department of Theoretical Physics, Faculty of Fundamental Problems of Technology, Wroc\l{}aw University of Science and Technology, 50-370 Wroc\l{}aw, Poland}

\author{Krzysztof Gawarecki}
	\affiliation{Department of Theoretical Physics, Faculty of Fundamental Problems of Technology, Wroc\l{}aw University of Science and Technology, 50-370 Wroc\l{}aw, Poland}

\author{Pawe{\l} Machnikowski}
	\affiliation{Department of Theoretical Physics, Faculty of Fundamental Problems of Technology, Wroc\l{}aw University of Science and Technology, 50-370 Wroc\l{}aw, Poland}
	
%\pacs{85.75.-d, 03.65.Yz}
%\date{\today}

\begin{abstract}
	We predict a spin pure dephasing channel in electron relaxation between states with unequal Zeeman splittings, exemplified by a spin-preserving electron tunneling between quantum dots in a magnetic field. The dephasing is caused by a mismatch in electron $g$-factors in the dots leading to distinguishability of phonons emitted during tunneling with opposite spins. Combining multiband $\kp$ modeling and dynamical simulations via a Master equation we show that this fundamental effect of spin measurement effected by the phonon bath may be widely controlled by the size and composition of the dots or on demand, via tuning of external fields. By comparing the numerically simulated degree of dephasing with the predictions of general theory based on distinguishability of environment states, we show that the proposed mechanism is the dominant phonon-related spin dephasing channel 
	and may limit spin coherence time in tunnel-coupled structures at cryogenic temperatures.
\end{abstract}

\maketitle

\section{Introduction}
	Quantum systems, distinguished from their classical counterparts by coherent superpositions of states, may lose this quantum nature via pure dephasing processes due to the build-up of correlations with the environment. The resulting emergence of classicality is connected with a transfer of the which-way information, the name of which comes from the analogy with double-slit interference experiments.\cite{FeynmanBook, SonnentagPRL2007} Demonstrations of such processes in various physical systems attract persistent attention.\cite{RoulleauPRL2009,FrabboniAPL2010,RichterPRL2015,YanagisawaSciRep2017}
	On the other hand, dephasing processes are critically destructive for coherent control of quantum states aimed at applications in spintronics\cite{KornPhysRep2010,ZuticRevModPhys2004, AwschalomBook2002,HansonRevModPhys2007} and quantum information processing,\cite{GalindoRevModPhys2002,GisinRevModPhys2002} where coherent superpositions of states, \textit{e.g.}, single spins in quantum dots (QDs),\cite{MikkelsenNP2007, RamsayPRL2008, GoddenAPL2010, GoddenPRB2012, GoddenPRL2012} can be used for computations beyond the classical schemes.\cite{Feynman1982,LossPRA1998}
	
	Recently, we have found in a simple model of carrier tunneling in a system of coupled QDs\cite{KrzykowskiAPPA2016} that spin-preserving orbital relaxation can be accompanied by spin pure dephasing if the environment response to the relaxation process depends on the spin state. This effect does not rely on any direct or spin-mixing-induced spin-environment interaction and is exclusively due to a misfit between electron $g$-factors, and hence Zeeman splittings, in the two QDs. The resulting unequal tunneling transition energies for the two spin states cause dephasing that may be interpreted in terms of which-way information: the state of the phonon bath after tunneling depends on the electron spin state. In this way, the bath ``measures'' the spin state via dissipated energy. Although this process takes place during tunneling (orbital relaxation), it does not lead to spin relaxation, and hence constitutes a pure dephasing on the spin subsystem. While spins in QDs offer long life times\cite{KroutvarNature2004} and, under certain conditions, coherence times,\cite{KuglerPRB2011} this dephasing channel may limit the fidelity of spin states in tunnel-coupled structures.\cite{MullerPRB2012}
	
	Here, we present accurate modeling of quantum states and dynamical simulations of spin pure dephasing that accompanies tunneling of electrons in double QD (DQD) systems. We focus on the controllability of the degree of dephasing: it depends on the spectral overlap of emitted phonon wave packets, which is determined by tunneling times and the Zeeman splittings mismatch only. We derive this fully general relationship explicitly from the Weisskopf-Wigner theory of spontaneous emission and make a quantitative connection between the distinguishability of phonons and the information about the spin state that leaks to the environment during tunneling.
	
	Comparing these calculations with the results of simulations containing all leading-order phonon-driven and spin-orbit effects, we determine the investigated channel to be the dominant phonon-related spin dephasing mechanism in the system under consideration. Its qualitative understanding and quantitative characterization allow us to propose ways of controlling the dephasing magnitude via appropriate sample design or tuning of external fields, the latter yielding feasible and promising methods of real-time control. This may lead to double-slit-like experiments with a continuously controlled level of ``observation'', but it may also be relevant for operations like spin-state transfer via non-resonant tunneling and spin-manipulation protocols in tunnel-coupled nanostructures. The dephasing mechanism itself is generic and should be manifested in transitions between states with different Zeeman splittings in any atomic, molecular or solid-state system.
	
	The paper is organized as follows. In Sec.~\ref{sec:system}, we define the system under study and present its theoretical modeling including multiband calculation of electron states and dynamical simulations. Next, in Sec.~\ref{sec:theory}, we derive the general expression for the spin coherence preserved after orbital relaxation. The results are presented in Sec.~\ref{sec:results}, where we also propose methods of dephasing control. Following this, in Sec.~\ref{sec:therm}, we investigate the additional dephasing that arises at finite temperatures. Finally, we conclude the paper in Sec.~\ref{sec:conclusions}. In the \hyperref[sec:appendix]{Appendix}, we give detailed information about numerically modeled structures.

\section{System and its theoretical modeling}\label{sec:system}
	We begin by describing the DQD system under investigation and methods used for its theoretical modeling. This is followed by a showcase of spin evolution for exemplary structures.

	\begin{table}[!tb]
		\caption{\label{tab:selected} Structural and calculated characteristics of selected structures at $T=\SI{0}{\kelvin}$.}
		\begin{ruledtabular}
		\begin{tabular}{l| d{0.1} d{0.5} | d{2.4} d{1.3} d{1.3} d{3.2} }
			\lab{$\mathrlap{\mkern-4mu\rule[-3.5pt]{\columnwidth}{0.25pt}}$Label}	&\lab{{$r_1$\,(nm)}}	&\lab{{$c_1$}}		&\lab{{$10^{3}\,\Delta g$}}	&\lab{{$\tau_+$\,(ns)}}	&\lab{{$\tau_-$\,(ns)}}	&\lab{{$10^{3}\,|\mathcal{C}|$}}	\\
			S3\Tstrut		& 12.5		& 0.5		& 94.7				& 1.778			& 1.601			& 7.12			\\
			S17			& 12.5		& 0.45		& 36.1				& 0.782			& 0.788			& 40.2			\\
			S32			& 13.4		& 0.4		& 5.38				& 0.807			& 0.806			& 232			\\
			S35			& 16.1		& 0.4		& 52.8				& 0.221			& 0.224			& 95.2			\\
			SX			& 12.5		& 0.4112		& 0.0097				& 2.043			& 2.042			& 499.97			\\
		\end{tabular}
		\end{ruledtabular}
	\end{table}
	\begin{figure}[!tb]
		\begin{center}
			\includegraphics[width=\columnwidth]{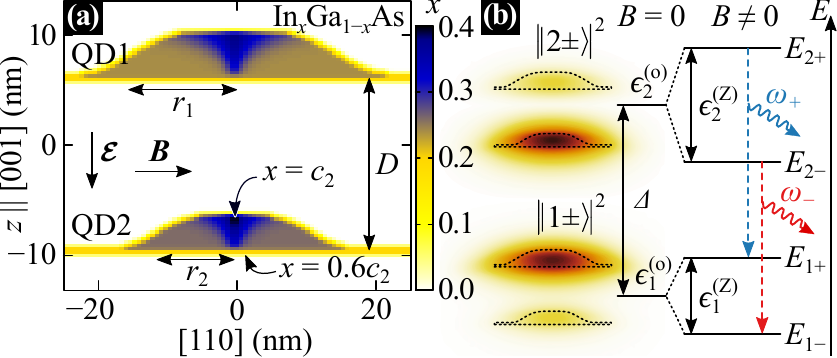}
		\end{center}
		\caption{\label{fig:fig1}(a) Exemplary structure material composition. (b) Projection of electron density in the two orbital states and the diagram of energy levels. Dashed arrows mark tunneling transitions.}
	\end{figure}
	We consider two vertically stacked, axially symmetric, coaxial, dome-shaped In$_x$Ga$_{1-x}$As QDs with a trumpet-shaped gradient of intradot In content\cite{MiglioratoPRB2002, JovanovPRB2012} (from $x\mathbin{=}0.6\,c_i$ at the base to $c_i$ at the top of QD$i$, $i\mathbin{=}1,2$), base radii $r_i$ and height $h_i\mathbin{=}r_i/3$, separated by the distance $D\mathbin{=}\SI{15.6}{\nano\meter}$ [see Fig.~\hyperref[fig:fig1]{\ref*{fig:fig1}(a)} for an exemplary material profile]. The structural parameters of QD2 are fixed: $r_2\mathbin{=}\SI{12}{\nano\meter}$ and $c_2\mathbin{=}0.43$, while for QD1 they are varied: $r_1\mathbin{=}\SIrange{10.7}{16.1}{\nano\meter}$ and $c_1\mathbin{=}0.35\mhyphen0.5$, which mainly alters the electron $g$-factor mismatch between QDs, $\Delta g$, but also, to a small extent, affects the phonon-assisted tunneling rates. In total, 50 structures were modeled to cover the $r_1\mhyphen c_1$ plane with a grid (see the \hyperref[sec:appendix]{Appendix}). In Table~\ref{tab:selected}, we present structural and calculated parameters of structures serving as exemplary throughout the paper, which are selected to represent various regimes of behavior.
	
	Electron wave functions are calculated within the 8-band envelope-function $\kp$ theory,\cite{BurtJPhysCondMat1992,ForemanPRB1993,BahderPRB1990} including spin-orbit effects, electric and magnetic fields,\cite{AndlauerPRB2008} strain,\cite{BirBOOK1974,PryorPRB1998} and piezoelectric field up to second-order terms\cite{BesterPRL2006,SchulzPRB2011,TseJAP2013} (see Ref.~[\onlinecite{GawareckiPRB2014}] for details of the model, numerical methods, and all material parameters except piezoelectric and spin-orbit coefficients that are taken from Refs.~[\onlinecite{CaroPRB2015}] and [\onlinecite{WinklerBOOK2003}], respectively). The axial electric field $\mathbcal{E}$ is used to tune the electron delocalization between QDs, hence the amount of occupation transferred during tunneling, to be the same for all structures. The basis of four lowest-energy states is computed, $\LR{\ket{{1-}}, \ket{{1+}}, \ket{{2-}}, \ket{{2+}}}$ with corresponding energies $E_{1(2)\pm}$, where $\ket{{\pm}} \mathbin{=} (\ket{\doa}\pm\ket{\upa}){/\!}\sqrt{2}$ are Zeeman eigenstates in the in-plane magnetic field $\bm{B}$ ($\ket{\upa/\doa}$ are spin eigenstates along the $z$-axis), and the number indicates the QD, in which the electron is mostly localized. The corresponding energy diagram and projections of probability densities in the two orbital states are shown in Fig.~\hyperref[fig:fig1]{\ref*{fig:fig1}(b)}. We shall speak of electron with spin ${\pm}$ in QD$i$, referring to the dominant spin and location of a given state.
	
	The acoustic-phonon bath is described with
	%
	%$H_{\mr{ph}}\mathbin{=}\sum_{\ql} \hbar \omega_{\ql} b^{\mathrlap{\dagger}}{}_{\ql} b_{\ql},$
	\[H_{\mr{ph}}=\sum_{\ql} \hbar \omega_{\ql} b^{\mathrlap{\dagger}}{}_{\ql} b_{\ql},\]
	where $b^{\mathrlap{\dagger}}{}_{\ql}$ creates a $\phbr$-branch phonon with wave vector $\phwv$, frequency $\omega_\ql\mathbin{=}q c_\phbr$, and velocity $c_\phbr$. The electron-phonon interaction enters via
	%
	%$H_\mr{int}\mathbin{=}\sumij \sigma_{ij} \cramped{\int} \mr{d}^3\rr \bm{\psi}_i^\dagger \lr{ H_\mr{B\mhyphen P}^{(\mr{ph})} + V_\mr{p}  } \,\bm{\psi}_j,$
	\begin{equation}
		H_\mr{int} =
			\sumij \sigma_{ij} \cramped{\int} \mr{d}^3\rr \, \bm{\psi}_i^\dagger\lr{\rr} \Lr[\big]{ H_\mr{B\mhyphen P}^{(\mr{ph})}\lr{\rr} + V_\mr{p}\lr{\rr}  } \,\bm{\psi}_j\lr{\rr},
	\end{equation}
	where $\sigma_{ij}\mathbin{=}\ketbra{i}{j}$, $\bm{\psi}_i$ is an 8-component pseudo-spinor of the $i$-th eigenstate envelope functions expressed in the standard $\kp$ basis,\cite{BahderPRB1990} $H_\mr{B\mhyphen P}^{(\mr{ph})}$ is the Bir-Pikus Hamiltonian evaluated with the phonon-induced strain field $\hat{\epsilon}_\mr{ph}$ to account for the deformation-potential coupling,\cite{WoodsPRB2004,RoszakPRB2007} 
	$V_\mr{p}\lr{\rr}\mathbin{=} \ii \lr{\hat{d}\hat{\epsilon}_\mr{ph}}_{\parallel} / {\varepsilon_0\varepsilon_\mr{r}}$
	is the phonon-induced piezoelectric field potential, and $\hat{d}$ is the third-rank piezoelectric tensor. While various higher-order phonon effects are known,\cite{KhaetskiiPRB2001,WoodsPRB2002,SanJosePRL2006} the leading-order contributions present in $H_{\bm{k}{\!\cdot}\bm{p}}\!+H_{\mr{int}}$ (including spin-orbit-\cite{KhaetskiiPRB2000,KhaetskiiPRB2001} and shear-strain-induced\cite{Mielnik-PyszczorskiPRB2018} admixture mechanisms, phonon-strain-driven: spin-orbit splitting of the electron spectrum and $g$-factor modification,\cite{KhaetskiiPRB2000,KhaetskiiPRB2001} the acoustic-phonon Pavlov-Firsov coupling,\cite{AlcaldePhysicaE2004,RomanoPRB2008,WangSSC2012} \textit{etc.}) tend to dominate under typical conditions.\cite{KhaetskiiPRB2001,WoodsPRB2002,Mielnik-PyszczorskiPRB2018}
	
	Orbital and spin degrees of freedom of the electron undergo dissipative evolution modeled with a non-secular Markovian Redfield equation\cite{BreuerBook2007} for the reduced density matrix,
	\begin{equation}\label{eq:redfield}
		\begin{aligned}[0.9\columnwidth]
			\dot{\rho} \lr{t} = {}&{}
				\frac{1}{\ii\hbar} \comm*{H_\mr{Z}}{\rho\lr{t}}
				+ \pi \sumijkl \LR[\Big]{ \e^{\ii \lr{ \widetilde{\omega}_{ij}-\widetilde{\omega}_{kl} } t} R_{jikl}\lr{\omega_{kl}} \\ %[-0.25em]
				{}&{} \times \Lr[\big]{ \sigma_{ kl} \rho\lr{t} \sigma_{ij}^\dagger - \sigma_{ij}^\dagger \sigma_{kl} \rho\lr{t} } + \hc},
		\end{aligned}
	\end{equation}
	written in the interaction picture with respect to the orbital energy
	%
	%${H_\mr{o} = \sum_{i} \epsilon_i^{(\mr{o})} \lr{ \ketbra{i+}{i+}+\ketbra{i-}{i-} }}$,
	\begin{equation} %\label{key}
		H_\mr{o} = \sum_{i} \epsilon_i^{(\mr{o})} \lr{ \ketbra*{i+}{i+}+\ketbra*{i-}{i-} },
	\end{equation}
	but with the Zeeman term
	%
	%${H_\mr{Z} = \sum_{i} g_i \mu_\mr{B} B \lr{ \ketbra*{i+}{i+}-\ketbra*{i-}{i-} } / 2}$
	\begin{equation} %\label{key}
		H_\mr{Z} = \hf \mu_\mr{B} B \sum_{i} g_i \lr{ \ketbra*{i+}{i+}-\ketbra*{i-}{i-} }
	\end{equation}
	kept in the Schr{\"o}dinger picture. Here,
	$g_i \mathbin{=} \epsilon_i^{(\mr{Z})}\!/\mu_\mr{B} B$
	are effective electron $g$-factors,
	$\epsilon_i^{(\mr{o})} \!= \lr{ E_{i+} \!+ E_{i-} } /2 $
	and
	$\epsilon_i^{(\mr{Z})} \!= E_{i+} \!- E_{i-}$
	are the respective orbital and Zeeman energy contributions,
	$\hbar\omega_{ij} = E_j\!-E_i,$
	$\hbar\widetilde{\omega}_{ij} = \epsilon_j^{(\mr{o}) } \!-\epsilon_i^{(\mr{o})},$
	%
	%$R_{ijkl} \lr{\omega}\mathbin{=} \cramped{\abs{ n \lr{-\omega} } \sum_\ql H_\mr{int}^{(ij)} H_\mr{int}^{(kl)} \delta \lr{ \abs{\omega} {-} \omega_\ql }/\hbar^2} \mathbin{=} R_{klij}\lr{\omega} \mathbin{=} R_{lkji}^{*}\lr{\omega}$
	\begin{align} %\label{key}
		R_{ijkl} \lr*{\omega}
			= {}&{} R_{klij}\lr*{\omega} = R_{lkji}^{*}\lr*{\omega} \nonumber \\
			= {}&{} \frac{1}{\hbar^{2}} \abs*{n \lr*{\omega}+1} \sum_\ql H_\mr{int}^{(ij)} H_\mr{int}^{(kl)} \, \delta \lr[\big]{ \abs*{\omega} - \omega_\ql } 
	\end{align}
	are phonon spectral densities,\cite{RoszakPRB2005}
	$H_\mr{int}^{(ij)} \!= \matrixel{i}{H_\mr{int}}{j},$
	and $n\lr{\omega}$ is the Bose distribution. Spin-dependent tunneling rates are determined from the Fermi's golden rule,
	%
	%$ \varGamma_{\pm}=\tau_\pm^{- 1}=2\pi R_{ijji} \lr{\omega_{ji}};\,\lr{i,j}\mathbin{=}\lr{{2\pm},{1\pm}}$.
	\begin{equation} %\label{key}
		\varGamma_{\pm} = \frac{1}{\tau_\pm} = 2\pi R_{ijji} \lr{\omega_{ji}};\quad \lr{i,j} = \lr{{2\pm},{1\pm}}.
	\end{equation}
	We solve Eq.~\eqref{eq:redfield} numerically with $\rho\lr{0}\mathbin{=}\ketbra{{2}{\upa}}{{2}{\upa}}$, corresponding to optical initialization with a circular polarization, for $B=\SI{5}{\tesla}$ and at $T=\SI{0}{\kelvin}$, unless otherwise stated.

	\begin{figure}[!tb]
		\begin{center}
			\includegraphics[width=\columnwidth]{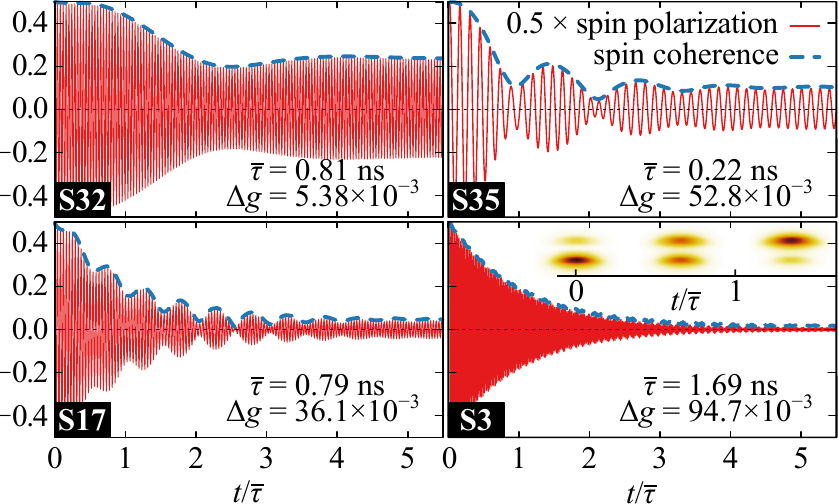}
		\end{center}
		\caption{\label{fig:fig2}Evolution of spin polarization (solid red lines) and coherence (dashed blue lines) for chosen structures. Time is given in units of average tunneling time $\widebar{\tau}$. Inset: evolution of electron density during tunneling.}
	\end{figure}
	In Fig.~\ref{fig:fig2}, we present the evolution of spin polarization $\sumi \lr{ \matrixel{i{\upa}}{\rho}{i{\upa}} - \matrixel{i{\doa}}{\rho}{i{\doa}} }$ and coherence $\abs{\sumi \matrixel{i{-}}{\rho}{i{+}}}$ during tunneling for selected structures differing mainly in the mismatch of Zeeman splittings between QDs, $\varDelta_\mr{Z}\mathbin{=}\Delta g \mu_\mr{B} B$. One may notice damping of spin precession accompanied by a proportional coherence loss, both related to but not uniquely determined by $\varDelta_\mr{Z}$, which suggests that another factor is involved. Importantly, the decoherence takes place once in the course of tunneling, and then the spin coherence becomes constant after a period of several spin-averaged tunneling times $\widebar{\tau}\mathbin{=}\lr{\tau_+ \!+ \tau_-}/2$.

\section{General theory}\label{sec:theory}
	To relate the above result to the discussed dephasing mechanism, we find the post-tunneling spin coherence within the Weisskopf-Wigner theory of spontaneous emission\cite{WeisskopfZfP1930,ScullyBOOK1997} adapted to the phonon bath and spin-dependent tunneling. Namely, we calculate spin coherence that would remain in the system if the only dephasing originated from the emission of distinguishable phonons.
	
	Spin-preserving tunneling is asymptotically described as
	\begin{equation}\label{eq:psi-tun-asymptot}
				\lr[\big]{ \alpha_{+} \ket{2+} + \alpha_{-} \ket{2-} } \ket{\varnothing}
				 \xrightarrow{\mkern-10mu t{\to}\infty\mkern-10mu} \alpha_{+} \ket{1{+}} \ket{\omega_{+}} + \alpha_{-} \ket{1{-}} \ket{\omega_{-}} ,
	\end{equation}
	where $\ket{\varnothing}$ is the bath initial state characterized by average phonon numbers $n_\ql$. Spin-dependent final states $\ket{\omega_{\pm}}$, corresponding to dissipated energies $\hbar\omega_\pm \!\mathbin{=} \varDelta \pm \varDelta_\mr{Z}/2$ [see Fig.~\hyperref[fig:fig1]{\ref*{fig:fig1}(b)}], are expanded in phonon modes,
	%
	%$\ket{\omega_\pm}\mathbin{=}\sum_{\ql} {c_\ql^{(\pm)}} b_{\ql}^{\dagger} \ket{\varnothing} / {\sqrt{n_\ql+1}};\,c_\ql^{(\pm)} \mathbin{\in} \mathbb{C}.$
	\begin{equation*}
		\ket{\omega_\pm}=
			\sum_{\ql} \frac{1}{\sqrt{n_\ql+1}}\, {c_\ql^{(\pm)}} \, b_{\ql}^{\dagger} \ket{\varnothing}\,; \quad c_\ql^{(\pm)} \in \mathbb{C}.
	\end{equation*}
	The spin-off-diagonal element of the target-QD (QD1) part of the reduced density matrix,
	%
	%$\rho\lr{\infty}\mathbin{=}\mr{Tr_{ph}} \lr{ \mathcal{C}_0 \ketbra{1{-}}{1{+}}\otimes\ketbra{\omega_{-}}{\omega_{+}}}+\dots;\,\mathcal{C}_0\mathbin{\equiv}\alpha_+^{*}\alpha_-$,
	\begin{equation} %\label{key}
		\rho\lr{\infty} = \mr{Tr_{ph}} \lr[\big]{ \mathcal{C}_0 \ketbra*{1{-}}{1{+}} \otimes \ketbra{\omega_{-}}{\omega_{+}} + \dots \,},
	\end{equation}
	where $\mathcal{C}_0\mathbin{\equiv}\alpha_+^{*}\alpha_-$, embodies the preserved spin coherence,
	\begin{equation}\label{eq:cdef}
		\mathcal{C}
			= \matrixel{1{-}}{\rho\lr{\infty}}{1{+}}
			= \mathcal{C}_0 \braket{\omega_{+}}{\omega_{-}}
			= \mathcal{C}_0 \sum_\ql {c_\ql^{({+})}}^{*} \, c_\ql^{({-})},
	\end{equation}
	strictly related to the overlap of bath states $\braket{\omega_{+}}{\omega_{-}}$, which is a measure of their distinguishability.\cite{DieksPhysLettA1988} We calculate the coefficients $c_\ql^{(\pm)}$, using $H_\mr{int}$ in the interaction picture and rotating-wave approximation,
	\begin{equation*}
		\mathcal{H}_\mr{int}
			= \hbar \sum_{\eta=\pm} \ketbra*{1\eta}{2\eta} \sum_{\ql} {g_\ql^{(\eta)}}^{*} \, b_\ql \,\e^{\ii \lr{ \omega_\eta - \omega_\ql } t} + \hc,
	\end{equation*} 
	where
	$g_\ql^{(\pm)} \mathbin{=} H_\mr{int}^{(1\pm2\pm)}\!/\hbar\mathbin{\equiv} g_\ql$
	for spin-diagonal coupling assumed here. We look for a solution to the Schr{\"o}dinger equation,
	$\ii\hbar\ket{\dot{\varPsi}\lr{t}}\mathbin{=}\mathcal{H}_\mr{int} \ket{\varPsi\lr{t}}$,
	in the form \eqref{eq:psi-tun-asymptot},
	\begin{align*}
		\ket{\varPsi \lr{t} } =
			{}&{} c\lr{t} \lr[\big]{ \alpha_{+} \ket{2{+}} + \alpha_{-} \ket{2{-}} } \otimes \ket{\varnothing} \\
			{}&{} +\sum_{\eta=\pm} \alpha_\eta \ket{1\eta} \otimes \sum_\ql \frac{1}{\sqrt{n_\ql+1}}\, c_\ql^{(\eta)} \lr{t} \,b_{\ql}^{\dagger} \ket{\varnothing}.
	\end{align*}
	This leads to a linear integro-differential equation for $c\lr{t}$,
	\begin{align*}
		\dot{c} \lr{t}
			= {}&{} -\sum_{\eta=\pm} \sum_\ql \abs[\big]{ g_\ql }^2 \int_0^t \mr{d}t' \, \e^{\ii \lr{ \omega_\eta-\omega_\ql }\lr{t-t'}} c\lr{t'} \\
			= {}&{} -\sum_{\eta=\pm} {\int_{-\infty}^{\infty}} \mr{d}\omega \, J\lr{\omega} {\int_0^t} \mr{d}t' \, \e^{\ii \lr{ \omega_\eta-\omega }\lr{t-t'}} c\lr{t'},
	\end{align*}
	where
	%
	%$J\lr{\omega}\mathbin{=}R_{2112}\lr{\omega}\vert_{T=\SI{0}{\kelvin}}\mathbin{=}\sum_{\ql} \abs{g_\ql}^2 \delta\lr{\omega-\omega_\ql}$.
	\[
		J\lr{\omega}
			= R_{2112}\lr{\omega} \big\vert_{T=\SI{0}{\kelvin}}
			= \sum_{\ql} \abs{g_\ql}^2 \, \delta\lr[\big]{\omega-\omega_\ql}.
	\]
	The leading contribution to the time integral comes from the $1{/}t\mathbin{\sim} \tau_\pm^{-1}$-wide vicinity of $\omega\mathbin{=}\omega_\pm$, where $J\lr{\omega} \mathbin{\approx} J\lr{\omega_\pm}\mathbin{=}\varGamma_{{\pm}}/2\pi$. Then, integration yields
	$c\lr{t} \mathbin{=}c\lr{0}\,\e^{-2\widebar{\varGamma} t}$,
	where
	$\widebar{\varGamma}\mathbin{=}\lr{\varGamma_{+} \!+ \varGamma_{-}}/2$. Next, the solution for $c_\ql^{(\pm)} \lr{t}$ is found,
	\begin{equation*}
		\cramped{c_\ql^{(\pm)} \lr{t}\mathbin{=}
			g_\ql \frac{1 \!- {\e^{\ii\lr{\omega_\ql{-}\omega_\pm}t-{\varGamma_{\pm}}t/2}}} {\omega_\ql-\omega_\pm + \mfrac{\ii \varGamma_{\pm}}{2}} \xrightarrow{\!\!{t\mathbin{\to}\infty}\!\!} \frac{g_\ql}{\omega_\ql \!- \omega_\pm \!+ \mfrac{\ii \varGamma_{\pm}}{2} }  }\,,
	\end{equation*}
	where the limit delivers the asymptotic $c_\ql^{(\pm)}$. Inserting the latter into Eq.~\eqref{eq:cdef}, we finally get
	\begin{equation}\label{eq:ww}
			\mathcal{C} = 
				\mathcal{C}_0\int_{-\infty}^{\infty} \frac{ \mr{d}\omega \, J\lr{\omega} } {\prod_{\eta{=}\pm} \lr{ \omega - \omega_{\eta} - \eta\mkern2mu \ii\varGamma_{{\eta}}/2 } }
				\simeq\frac{\mathcal{C}_0}{ 1 - \ii \varDelta_\mr{Z} / {\hbar \widebar{\varGamma}}},
	\end{equation}
	where $\mathcal{C}_0\mathbin{=}1/2$ in our case and the integral is calculated with the residue theorem, assuming that $\varGamma_{{\pm}}\mathbin{\ll}\omega_{\pm}$.
	
	According to Eq.~\eqref{eq:ww}, faster tunneling is favorable for higher coherence, but it is not a matter of competition between rates of tunneling and decoherence. The dephasing is a one-time process tied to the tunneling and it is not characterized by own rate. Instead, the characteristic parameter is $\varDelta_\mr{Z}$, which is not related directly to any rate and defines the degree of coherence loss, rather than its time scale.
	
\section{Results and dephasing control protocols}\label{sec:results}	
	In this Section, we compare the numerical results from Sec.~\ref{sec:system} with those obtained according to the general theory from Sec.~\ref{sec:theory}. Then, we propose methods for controlling the degree of dephasing.

	\begin{figure}[!tb]
		\begin{center}
			\includegraphics[width=\columnwidth]{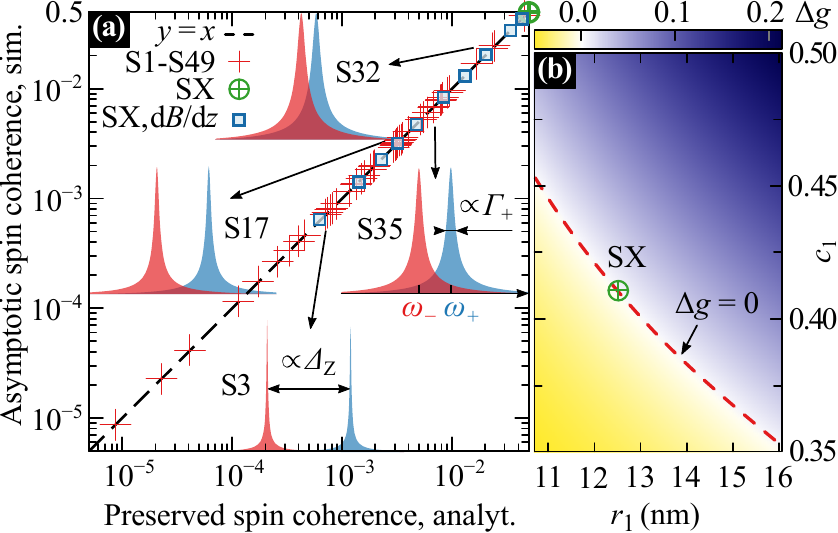}
		\end{center}
		\caption{\label{fig:fig3}(a) Spin coherence preserved after tunneling: simulation, Eq.~\protect\eqref{eq:redfield}, vs.\ analytical solution, Eq.~\protect\eqref{eq:ww}, for structures S1-S49 ({\large {\color{cbred} +}}), SX ({{\color{cbgreen}$\raisemath{0.5pt}{\bm{\oplus}}$}}) and for SX with a magnetic-field gradient ({\large{\color{cbblue} $\Box$}}). Insets: spectra of phonon wave packets emitted during tunneling in the two spin states for chosen structures. (b) Interpolated dependence of the $g$-factor mismatch $\Delta g$ on QD1 size $r_1$ and In content $c_1$.}
	\end{figure}
	In Fig.~\hyperref[fig:fig3]{\ref*{fig:fig3}(a)}, we confront preserved spin coherence $\mathcal{C}$ calculated according to Eq.~\eqref{eq:ww} with post-tunneling values obtained from numerical simulations for 49 structures. The two match perfectly, proving that the discussed dephasing is the dominant spin decoherence channel in the system, apart from possible higher-order phonon couplings beyond our model.\cite{SanJosePRL2006} Moreover, by varying the QD1 size and composition within reasonable ranges, one may cover the full range of $\abs{\mathcal{C}}$ values. We propose to use this tunability to design structures of desired properties. While, according to Eq.~\eqref{eq:ww}, the preserved coherence is a function of $\varDelta_\mr{Z}$ and $\varGamma_\pm$, these parameters depend on the QD morphology, as well as on external fields. Fig.~\hyperref[fig:fig3]{\ref*{fig:fig3}(b)} shows the dependence of $\Delta g$ on the size and composition of QD1, indicating that it varies considerably and changes sign when the morphology is modified within a relatively narrow range of realistic values. The green circles in Figs.~\hyperref[fig:fig3]{\ref*{fig:fig3}(a)} and \hyperref[fig:fig3]{\ref*{fig:fig3}(b)} correspond to an additional structure SX, designed to cancel the mismatch of Zeeman splittings, and hence the dephasing.
	
	Let us note that the effect is expected in any material system: the variation of $g$-factor with QD size is unavoidable and bulk $g$-factors\cite{RothPhysRev1959,WillatzenPRB1995} for common systems exhibit substantial mismatches, often higher that for InGaAs discussed here.
	Additionally, $\varDelta_\mr{Z}\mathbin{\neq}0$ may also arise, even for $\Delta g\mathbin{=}0$, due to a magnetic-field gradient. Blue squares in Fig.~\hyperref[fig:fig3]{\ref*{fig:fig3}(a)} correspond to the structure SX, which has negligible $g$-factor mismatch, subject to field gradients in the range $0.65\mathbin{\times}10^{-n/3}\,\si{\tesla\per\nano\metre}$, $n=0,\dots,9$ (from left to right). The resulting dephasing is equivalent to that caused by $\Delta g$. For the highest simulated gradient (orders of magnitude higher than $\SIrange{\sim 0.1}{1}{\milli\tesla/\nano\metre}$ met in current nanodevices\cite{YonedaAPX2015}), a \SI{6}{\percent} deviation from Eq.~\eqref{eq:ww} occurs, resulting from enhanced spin-flips in the high local field.

	\begin{figure}[!tb]
		\begin{center}
			\includegraphics[width=\columnwidth]{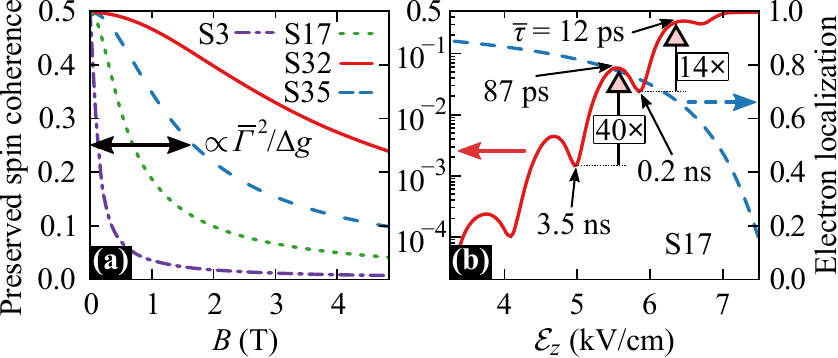}
		\end{center}
		\caption{\label{fig:fig4}(a) Dependence of spin coherence preserved after tunneling $\abs{\mathcal{C}}$ on magnetic field $B$ for chosen structures. (b) Dependence of $\abs{\mathcal{C}}$ (solid red line, left axis) and electron localization (dashed blue line, right axis) on axial electric field $\mathcal{E}_z$. Vertical arrows mark possible coherence gains. Tunneling times are given at local extrema.}
	\end{figure}
	A practical control protocol should rely on external fields applied to the sample rather than manufacturing conditions. Obviously, since dephasing is due to the mismatch of Zeeman splittings, it can be eliminated by reducing the magnitude of the magnetic field. The dependence of $\abs{\mathcal{C}}$ on the latter is plotted for chosen structures in Fig.~\hyperref[fig:fig4]{\ref*{fig:fig4}(a)}, showing a Lorentzian-like shape of width $\mathbin{\propto}\widebar{\varGamma}^2\!/ \Delta g$. Thus, a fast tunneling regime (represented by S32) may be used to widen the range of $B$ for reasonably coherent tunneling. On the other hand, slow tunneling allows one to toggle coherence on and off with small changes in low magnetic field, \textit{e.g.}, $\abs{\mathcal{C}}$ from 0.035 at $B\mathbin{=}\SI{1}{\tesla}$ to 0.5 at $B\mathbin{=}\SI{0}{\tesla}$ for structure S3. In principle, a gradient $\mr{d}B_x/\mr{d}z\mathbin{\simeq} -2 B_0 \Delta g /D\lr{g_1+g_2}$ could be used to cancel $\varDelta_\mr{Z}$, although this may be hardly feasible.
	
	Another way of controlling spin dephasing is to use oscillations in tunneling rates with transition energy (period $\mathbin{\propto}D^{-1}$)\cite{WijesundaraPRB2011,GawareckiPRB2010} that may be tuned with an axial electric field. This changes also the degree of electron localization, up to now kept fixed. Both these dependencies are plotted in Fig.~\hyperref[fig:fig4]{\ref*{fig:fig4}(b)}. We find that decoherence can be controlled within a range of values extending over many orders of magnitude, while keeping the electron localized and with reasonable tunneling times. Thus, a feasible, on-demand control of dephasing is possible.

\section{Cumulative dephasing at finite temperature}\label{sec:therm}
	While in the previous sections we dealt with the $T\mathbin{=}\SI{0}{\kelvin}$ limit, here we study the additional dephasing that arises at finite temperatures.

	\begin{figure}[!tb]
		\begin{center}
	 		\includegraphics[width=\columnwidth]{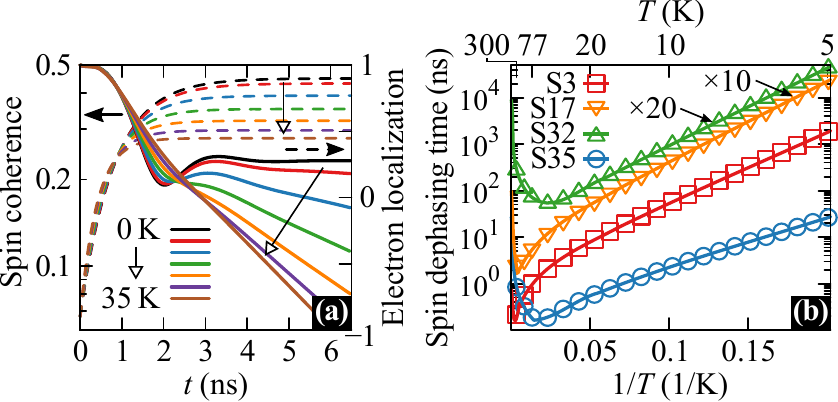}
		\end{center}
		\caption{\label{fig:fig5}(a) Evolution of spin coherence (solid lines, left axis) and electron localization (dashed lines, right axis) for structure S32 at various temperatures. (b) Arrhenius plot of post-tunneling spin dephasing times from simulations (symbols) and calculated with Eq.~\protect\eqref{eq:gamma_d} (lines) for selected structures.}
	\end{figure}
	Saturation of spin coherence in Fig.~\ref{fig:fig2} results from the fact that tunneling at $T\mathbin{=}\SI{0}{\kelvin}$ is irreversible, hence the dephasing takes place once. At $T\mathbin{>}\SI{0}{\kelvin}$, thermally activated back-tunneling enables a continuous dephasing process, accompanying the repeated virtual tunneling between the QDs. For a better insight, we analyze the Master equation \eqref{eq:redfield} analytically in the interaction picture with respect to $H_\mr{Z}$. Labels are here changed for brevity: $\LR{1-,1+,2-,2+}\to\LR{\oned,\oneu,\twod,\twou}$. To account only for the studied mechanism, we neglect all spin-off-diagonal couplings by setting $R_{ijkl}\mathbin{=}0$ if either of pairs $(i,j)$, $(k,l)$ is spin-mismatched and assume couplings to be spin-invariant, \textit{i.e.}, $R_{ij\oneu\twou}\mathbin{=}R_{ij\oned\twod}$, \textit{etc}. This decouples equations for spin coherence,
	\begin{subequations}\label{eq:eqns}
		\begin{align}
			&\begin{aligned}\label{eq:eqnsa}
			\matrixel{\twod}{\dot{\rho}}{\twou} \simeq {}&{}
				-\pi \Lr[\big]{ \mkern5.75mu R\lr{ \omega_{+}} \mkern5.75mu +\mkern5.75mu  R\lr{ \omega_{-}} \mkern5.75mu } \matrixel{\twod}{{\rho_{}}}{\twou} \\
				{}&{} +\pi \Lr[\big]{ R\lr{-\omega_{+}} + R\lr{-\omega_{-}} } \matrixel{\oned}{{\rho_{}}}{\oneu} \,\e^{-\ii \omega_\mr{Z} t},
			\end{aligned}\\
			&\matrixel{\oned}{\dot{\rho}}{\oneu} \simeq \,-\, \matrixel{\twod}{\dot{\rho}}{\twou} \,\e^{\ii \omega_\mr{Z} t},\label{eq:eqnsb}
		\end{align}
	\end{subequations}
	where $R\lr{\omega}\mathbin{\equiv}R_{2112}\lr{\omega}$,
	and $\omega_{\mr{Z}} \mathbin{\equiv} \varDelta_\mr{Z}/{\hbar} \mathbin{=} {\omega}_{+}\!-{\omega}_{-}$.
	At $T\mathbin{=}\SI{0}{\kelvin}$, $R\lr{-\omega_\pm}$ vanish and only the first term in Eq.~\eqref{eq:eqnsa} is left, describing an exponential outflow of coherence from QD2 at the rate 
	$\pi \Lr{ R\lr{ \omega_{+}} + R\lr{ \omega_{-}} }\mathbin{=}{\widebar{\varGamma}}$,
	\textit{i.e.}, exactly during tunneling. The associated inflow to QD1, according to Eq.~\eqref{eq:eqnsb}, is affected by a phase factor that oscillates with a frequency equal to the mismatch of Larmor precession frequencies $\omega_\mr{Z}$. The solution for the target-QD spin coherence,
	\begin{equation*}
		\mathcal{C}(t) \simeq
			\matrixel{{\oned}}{\rho}{{\oneu}}\vert_{T=\SI{0}{\kelvin}}=
			\mathcal{C}_0 \frac{1-\e^{\ii \omega_\mr{Z} t - \widebar{\varGamma}t}}{1-\ii {\omega_\mr{Z}}/{\widebar{\varGamma} } }\xrightarrow{t\to\infty} \mathcal{C},
	\end{equation*}
	reproduces Eq.~\eqref{eq:ww} derived within the spontaneous emission theory.
	The terms that arise in Eqs.~\eqref{eq:eqns} at $T\mathbin{>}\SI{0}{\kelvin}$ describe the transfer of spin coherence in the opposite direction via thermally activated back-tunneling at a rate
	%
	%$\varGamma_{\mr{v}} \mathbin{\equiv} \lr{ \e^{{-}{\hbar\omega_{+}}/{k_{}T}} \varGamma_{+} \!+  \e^{{-}{\hbar\omega_{-}}/{k_{}T}}  \varGamma_{-} }/2 \mathbin{\simeq} \widebar{\varGamma} \,\e^{{-}{\Delta}/{k_{}T}}$.
	\[
		\varGamma_{\mr{v}}
			\equiv \hf \lr[\big]{ \e^{{-}{\hbar\omega_{+}}/{k_{}T}} \varGamma_{+} \!+  \e^{{-}{\hbar\omega_{-}}/{k_{}T}}  \varGamma_{-} }
			\simeq \widebar{\varGamma} \,\e^{{-}{\varDelta}/{k_{}T}}.
	\]
	This leads to the solution with a long-time exponential decay of spin coherence at the rate given by
	\begin{equation}\label{eq:gamma_d}
		2\varGamma_{\mr{d}} =
			\widebar{\varGamma} + \varGamma_{\mr{v}} - \Re \sqrt{\lr{\widebar{\varGamma} + \varGamma_{\mr{v}}}^2 + 2\ii  \omega_{\mr{Z}} \lr{ \varGamma_{\mr{v}} \!- \widebar{\varGamma} } -\omega_{\mr{Z}}^{2} } \,.
	\end{equation}
	Thus, at $T\mathbin{>}\SI{0}{\kelvin}$, after the dephasing during tunneling, the remaining coherence undergoes exponential decay of similar origin. Temperature-driven emergence of the latter in the equilibrated part of evolution is visible in Fig.~\hyperref[fig:fig5]{\ref*{fig:fig5}(a)}, where the simulated spin coherence is presented along with electron localization.
	The dependence of corresponding dephasing times on temperature is shown in Fig.~\hyperref[fig:fig5]{\ref*{fig:fig5}(b)}, where values extracted from numerical simulations (symbols) are compared to $\varGamma_\mr{d}$ calculated according to Eq.~\eqref{eq:gamma_d} (lines). The agreement indicates that, up to room temperature, the considered dephasing channel dominates over other processes included in numerical simulations but deliberately neglected in the analytical solution for $\varGamma_\mr{d}$. Interestingly, a non-monotonic behavior is present due to the interplay between the temperature-driven rise of accumulative dephasing and the overall decrease of phonon distinguishability with rising tunneling rates.
	
	While the one-time dephasing applies to carriers that actually tunnel, the exponential decoherence at $T\mathbin{>}\SI{0}{\kelvin}$ may affect also stationary spins in tunnel-coupled structures.
	The dephasing time depends nontrivially on system parameters and widely varies among simulated structures. For $B 
	\mathbin{=} \SI{5}{\tesla}$ it may be as short as \SI{\sim10}{\nano\second} at $T \mathbin{=} \SI{5}{\kelvin}$ and \SI{\sim0.1}{\nano\second} at $T \mathbin{=} \SI{77}{\kelvin}$. Hence, it may vary from comparable to a few orders of magnitude shorter than homogeneous dephasing times due to hyperfine interaction\cite{PettaPhysE2006,BechtoldNatPhys2015,StockillNatCommun2016} (\si{\micro\second} range) and, under certain conditions, may even surpass the inhomogeneous dephasing ($\SIrange{\sim2}{20}{\nano\second}$\cite{PettaPhysE2006,BechtoldNatPhys2015,StockillNatCommun2016}). The other relevant source of spin perturbation, the charge noise, has been recently found to lead to slower spin pure relaxation.\cite{HuangPRB2014} Thus, the discussed effect may be the one to limit spin coherence time in tunnel-coupled structures.

\section{Conclusions}\label{sec:conclusions}
	We have presented a prediction of a spin pure dephasing channel in a spin-preserving electron orbital relaxation in a magnetic field between states with unequal Zeeman splittings. The dephasing originates from distinguishability of reservoir excitations induced by the mismatch of energies dissipated during transitions in the two spin states, and hence the resulting leakage of information on the spin superposition to the environment. The mechanism is thus general and of fundamental nature analogous to measuring the position of a particle in double-slit experiments. Our theoretical analysis of reservoir state distinguishability within the theory of spontaneous emission showed that the effect depends on the difference of Zeeman splittings and relaxation rates. These parameters define the spectral overlap of phonon wave packets emitted during relaxation in the two spin states. Additionally, we have presented a detailed quantitative analysis of spin dephasing for the case of spin-preserving tunneling between self-assembled QDs via a realistic multi-band $\kp$ modeling and simulations using a non-secular Markovian master equation. In this case, the effect is expected in virtually any material system. We have shown that the considered mechanism is the only leading-order phonon-related spin perturbation relevant in the double QD system and that it may limit spin coherence time in tunnel-coupled structures at cryogenic temperatures. Finally, we have proposed ways of controlling spin decoherence both at the stage of sample manufacturing via the size and composition of QDs, and on demand, by tuning of external fields. The latter promises a feasible method of a real-time control over spin decoherence in a range that covers many orders of magnitude.

\begin{acknowledgments}
	We acknowledge support from the Polish National Science Centre by Grants No.~{2014/\allowbreak{}13/\allowbreak{}B/\allowbreak{}ST3/\allowbreak{}04603} (K.G., P.M.) and No.~{2011/\allowbreak{}02/\allowbreak{}A/\allowbreak{}ST3/\allowbreak{}00152} (M.G.). We would also like to thank M.~Syperek for an inspiring discussion. Numerical calculations have been carried out using resources provided by Wroclaw Centre for Networking and Supercomputing (\url{http://wcss.pl}), Grant No.~203.
\end{acknowledgments}

\appendix*
\renewcommand\theequation{A\arabic{equation}}
\renewcommand\thefigure{A\arabic{figure}}
\renewcommand\thetable{A\Roman{table}}
\setcounter{equation}{0}
\setcounter{table}{0}
\setcounter{figure}{0}
\section{Details of numerically modeled structures}\label{sec:appendix}
	In this Appendix, we describe the set of numerically modeled DQD structures used in the paper, give their structural and calculated characteristics, as well as describe the procedure used for interpolation of $\Delta g$ values.
	
	\begingroup \squeezetable
	\begin{table}[tb]
		\newcommand{\highlightt}[1]{\raisebox{1.pt}{\makebox[0pt][l]{\fboxsep-1pt\hspace{-0pt}\colorbox{#1!7} {\strut\hspace*{0.99\linewidth}}}}\!\!}
		\begin{tabular*}{\textwidth}{l|@{\extracolsep{\fill}} S[table-format=0.1,detect-mode] S[table-format=0.4,detect-mode] S[table-format=2.2,detect-mode] S[table-format=0.2,detect-mode] S[table-format=0.2,detect-mode] S[table-format=2.3,detect-mode] S[table-format=2.3,detect-mode] S[table-format=3.4,detect-mode]}
			\hline\hline\rule{0pt}{1.15em}
				& {${r_1}$}	& {$c_1$} & {$\Delta g$} 	& {$\hbar\omega_+$}	& {$\hbar\omega_-$}	& {$\tau_+$}	& {$\tau_-$}	& {$\abs{\mathcal{C}}$}	\\
			\rule{0pt}{1.25em}	& {(nm)} 		& 		& {$\times 10^{3}$} & {(meV)}			& {(meV)} 		& {(ns)}		& {(ns)}		& {$\times 10^{3}$}		\\
			\hline \rule{0pt}{1.15em}\!\!
			S1	& 10.7	&	0.5		&	41.0		&	4.41	&	4.40	&	10.8		&	10.5		&	2.61		\\
			S2	& 11.6	&	0.5		&	67.2		&	3.61	&	3.60	&	0.851	&	0.886	&	19.5		\\	\highlightt{RoyalBlue}
			S3	& 12.5	&	0.5		&	94.7		&	3.02	&	2.99	&	1.78		&	1.60		&	7.12		\\
			S4	& 13.4	&	0.5		&	123		&	2.52	&	2.49	&	0.527	&	0.636	&	15.9		\\
			S5	& 14.3	&	0.5		&	153		&	2.13	&	2.08	&	0.226	&	0.237	&	32.1		\\
			S6	& 15.2	&	0.5		&	183		&	1.81	&	1.75	&	0.329	&	0.301	&	19.8		\\
			S7	& 16.1	&	0.5		&	213		&	1.52	&	1.46	&	0.601	&	0.607	&	8.87		\\
			S8	& 10.7	&	0.475	&	17.8		&	4.70	&	4.69	&	5.53		&	5.50		&	11.6		\\
			S9	& 11.6	&	0.475	&	40.1		&	3.88	&	3.87	&	2.05		&	2.17		&	13.4		\\
			S10	& 12.5	&	0.475	&	63.7		&	3.21	&	3.20	&	0.921	&	0.892	&	19.7		\\
			S11	& 13.4	&	0.475	&	88.6		&	2.70	&	2.67	&	1.51		&	1.80		&	7.78		\\
			S12	& 14.3	&	0.475	&	114		&	2.28	&	2.24	&	0.269	&	0.294	&	35.4		\\
			S13	& 15.2	&	0.475	&	140		&	1.96	&	1.92	&	0.243	&	0.241	&	33.4		\\
			S14	& 16.1	&	0.475	&	167		&	1.65	&	1.60	&	0.48		&	0.431	&	15.0		\\
			S15	& 10.7	&	0.45		&	-2.21	&	4.97	&	4.97	&	6.15		&	6.14		&	82.6		\\
			S16	& 11.6	&	0.45		&	16.2		&	4.09	&	4.09	&	8.49		&	8.80		&	8.13		\\	\highlightt{RoyalBlue}
			S17	& 12.5	&	0.45		&	36.1		&	3.43	&	3.42	&	0.782	&	0.788	&	40.2		\\
			S18	& 13.4	&	0.45		&	57.2		&	2.88	&	2.86	&	2.77		&	2.65		&	7.36		\\
			S19	& 14.3	&	0.45		&	79.2		&	2.44	&	2.41	&	0.445	&	0.496	&	30.5		\\
			S20	& 15.2	&	0.45		&	101		&	1.99	&	1.96	&	0.214	&	0.214	&	52.2		\\
			S21	& 16.1	&	0.45		&	125		&	1.77	&	1.74	&	0.320	&	0.301	&	29.4		\\
			S22	& 10.7	&	0.425	&	-19.1	&	5.24	&	5.24	&	13.1		&	12.8		&	4.59		\\
			S23	& 11.6	&	0.425	&	-4.39	&	4.37	&	4.37	&	14.2		&	14.2		&	18.2		\\
			S24	& 12.5	&	0.425	&	11.8		&	3.66	&	3.66	&	1.19		&	1.20		&	79.3		\\
			S25	& 13.4	&	0.425	&	29.4		&	3.06	&	3.05	&	1.26		&	1.23		&	31.0		\\
			S26	& 14.3	&	0.425	&	48.0		&	2.61	&	2.59	&	1.16		&	1.28		&	19.4		\\
			S27	& 15.2	&	0.425	&	67.0		&	2.20	&	2.18	&	0.245	&	0.256	&	67.3		\\
			S28	& 16.1	&	0.425	&	86.8		&	1.90	&	1.88	&	0.239	&	0.236	&	54.8		\\
			S29	& 10.7	&	0.4		&	-32.9	&	5.56	&	5.57	&	86.9		&	80.9		&	0.413	\\
			S30	& 11.6	&	0.4		&	-21.6	&	4.62	&	4.63	&	6.49		&	6.54		&	8.08		\\
			S31	& 12.5	&	0.4		&	-8.83	&	3.90	&	3.90	&	3.79		&	3.72		&	34.3		\\	\highlightt{RoyalBlue}
			S32	& 13.4	&	0.4		&	5.38		&	3.30	&	3.30	&	0.807	&	0.806	&	232		\\
			S33	& 14.3	&	0.4		&	20.5		&	2.80	&	2.79	&	3.05		&	3.04		&	18.2		\\	
			S34	& 15.2	&	0.4		&	36.3		&	2.39	&	2.38	&	0.444	&	0.467	&	68.2		\\	\highlightt{RoyalBlue}
			S35	& 16.1	&	0.4		&	52.8		&	2.05	&	2.03	&	0.221	&	0.224	&	95.2		\\
			S36	& 10.7	&	0.375	&	-43.4	&	5.85	&	5.87	&	295		&	311		&	0.0865	\\
			S37	& 11.6	&	0.375	&	-35.5	&	4.92	&	4.93	&	8.00		&	7.82		&	4.05		\\
			S38	& 12.5	&	0.375	&	-25.9	&	4.15	&	4.16	&	25.8		&	25.3		&	1.72		\\
			S39	& 13.4	&	0.375	&	-14.9	&	3.50	&	3.51	&	0.967	&	0.957	&	78.5		\\
			S40	& 14.3	&	0.375	&	-2.91	&	2.99	&	2.99	&	1.39		&	1.40		&	244		\\
			S41	& 15.2	&	0.375	&	9.76		&	2.57	&	2.57	&	1.25		&	1.27		&	91.1		\\
			S42	& 16.1	&	0.375	&	23.2		&	2.21	&	2.20	&	0.275	&	0.280	&	167		\\
			S43	& 10.7	&	0.35		&	-50.8	&	6.17	&	6.18	&	97.5		&	99.9		&	0.227	\\
			S44	& 11.6	&	0.35		&	-46.0	&	5.20	&	5.21	&	22.6		&	21.1		&	1.13		\\
			S45	& 12.5	&	0.35		&	-39.4	&	4.44	&	4.45	&	8.62		&	8.89		&	3.30		\\
			S46	& 13.4	&	0.35		&	-31.4	&	3.77	&	3.78	&	2.75		&	2.60		&	13.6		\\
			S47	& 14.3	&	0.35		&	-22.3	&	3.21	&	3.21	&	0.835	&	0.837	&	60.5		\\
			S48	& 15.2	&	0.35		&	-12.6	&	2.75	&	2.76	&	2.89		&	2.89		&	31.3		\\
			S49	& 16.1	&	0.35		&	-2.07	&	2.36	&	2.36	&	0.466	&	0.464	&	460		\\	\highlightt{red} 
			SX	& 12.5	&	0.4112	&	0.0097	&	3.72	&	3.72	&	2.04		&	2.04		&	499.97	\\
			\hline\hline
		\end{tabular*}
		\caption{\label{tab:all}Characteristics of modeled structures: QD1 radius $r_1$ and In content $c_1$, calculated $g$-factor mismatch $\Delta g$, transition energies $\hbar\omega_\pm$ and tunneling times $\tau_\pm$ for electrons in the two spin states, as well as spin coherence preserved after tunneling $\abs{\mathcal{C}}$ at $T\mathbin{=}\SI{0}{\kelvin}$ and $B\mathbin{=}\SI{5}{\tesla}$. Structures used as exemplary in the paper are highlighted.}
	\end{table}\endgroup
	\begin{figure}[tb]
		\begin{center}
			\includegraphics[width=\columnwidth]{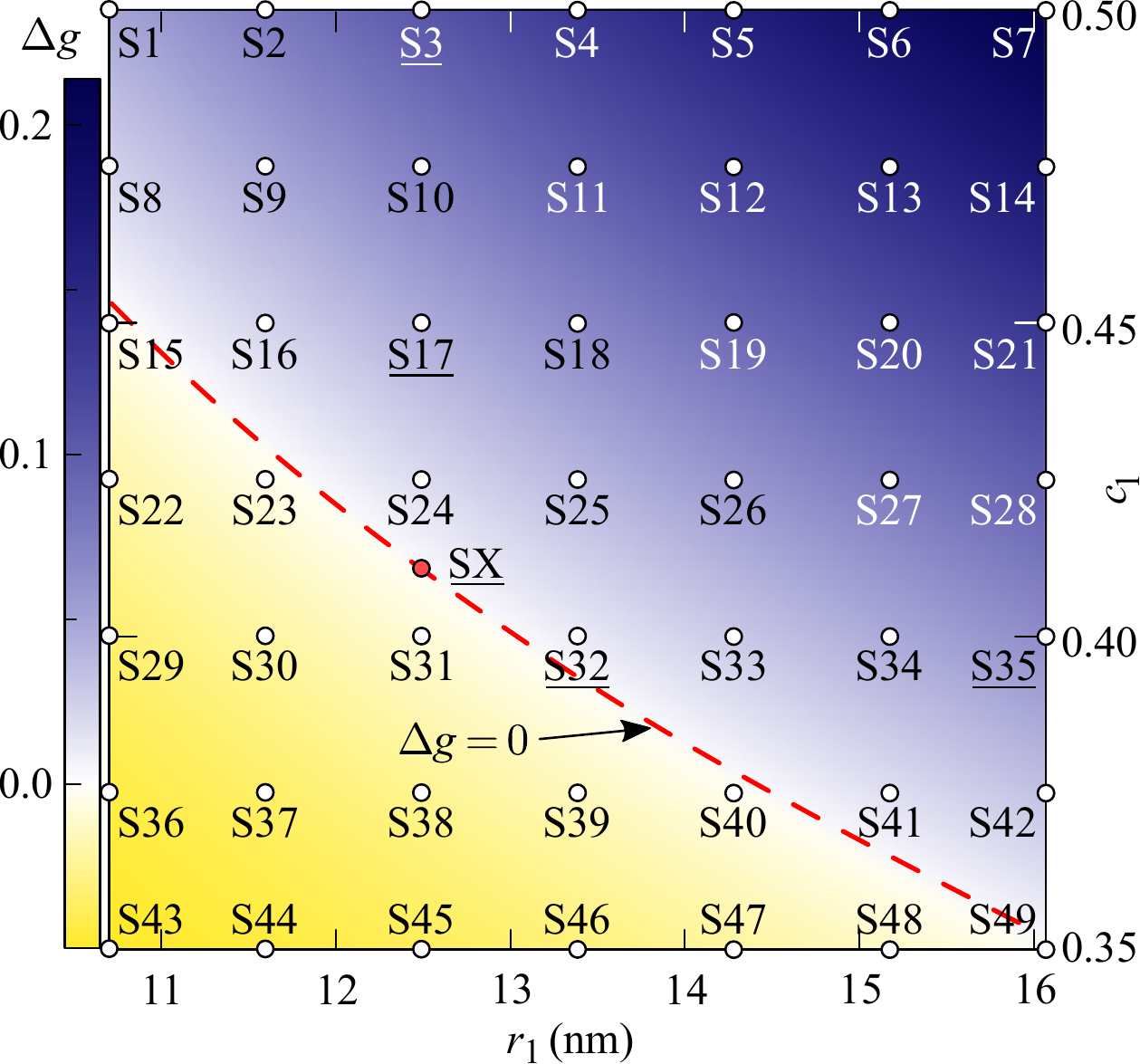}
		\end{center}
		\caption{\label{fig:supl_fig}Interpolated dependence of the $g$-factor mismatch $\Delta g$ on QD1 size $r_1$ and In content $c_1$. Parameters (size and composition of QD1) of structures are represented by position of dots with their labels, underlined for structures used as exemplary in the paper.}
	\end{figure}
	In total, we modeled 50 DQD structures of realistically varying morphology. Those labeled S1-S49 differ in the QD1 base radius $r_1$ and indium composition $c_1$ [see Fig.~\hyperref[fig:fig1]{\ref*{fig:fig1}(a)}] and form a $7\mathbin{\times}7$ axis-wise regular grid in the $r_1\mhyphen c_1$ plane. The additional structure SX is designed to simulate a dephasing-free system with approximately equal $g$-factors in the two QDs. Characteristics of all modeled structures are presented in Table~\ref{tab:all}, where also calculated $g$-factor mismatch $\Delta g$, tunneling times $\tau_{\pm}$ and transition energies $\hbar\omega_{\pm}$ as well as spin coherence preserved after tunneling, $\abs{\mathcal{C}}$, at $T\mathbin{=}\SI{0}{\kelvin}$ and $B\mathbin{=}\SI{5}{\kelvin}$ are given. In Fig.~\ref{fig:supl_fig}, the structures are represented by positions of dots with respective labels in the $r_1\mhyphen c_1$ plane, where also the interpolated dependence of $\Delta g$ is plotted as a color map. The interpolation was done with a surface of the first and second order in $r_1$ and $c_1$, respectively. This is justified, as we expect a close to linear dependence of electron $g$-factor on a uniform QD size change\cite{NakaokaPRB2004,vanBreePRB2012} and terms up to second order in the In concentration, presumably inherited after quadratic corrections to the interpolation of the bulk $g$-factor between GaAs and InAs, which is commonly referred to as \textit{bowing}.\cite{VurgaftmanJAP2001}
\FloatBarrier

%\bibliography{whichway}
%

\end{document}